\documentclass[journal]{IEEEtran}
\usepackage{ifpdf}
\ifpdf
    \usepackage[pdftex]{graphicx}
    \graphicspath{{figs/}{matlab/}}   
    \usepackage[update]{epstopdf}
\else
	\usepackage{graphicx}
	\graphicspath{{figs/}{matlab/}}   
\fi

\usepackage{array}
\usepackage{amsmath}
\usepackage{amsfonts}
\usepackage{amssymb}
\usepackage{amstext}
\usepackage{latexsym}
\usepackage{color}
\usepackage{cite,url}
\usepackage{setspace}
\newtheorem{thm}{Theorem}
\newtheorem{prop}{Proposition}
\newtheorem{lemma}{Lemma}

\newtheorem{coro}[lemma]{Corollary}

\usepackage{pdflscape}
\allowdisplaybreaks
\usepackage{multirow}

\DeclareMathSizes{10}{9}{8}{7}

\begin{document}
\title{\huge New Results on Multilevel Diversity Coding with Secure Regeneration}
\author{Shuo Shao,~\IEEEmembership{Student Member,~IEEE}, Tie Liu,~\IEEEmembership{Senior Member,~IEEE}, \\Chao Tian,~\IEEEmembership{Senior Member,~IEEE} and Cong Shen,~\IEEEmembership{Senior Member,~IEEE}
\thanks{S. Shao is with the Department of Electrical Engineering, Shanghai Jiaotong University, Shanghai, China (e-mail:shuoshao@sjtu.edu.cn)}
\thanks{T.~Liu and C.~Tian are with the Department of Electrical and Computer Engineering, Texas A\&M University, College Station, TX 77843, USA (e-mail: \{tieliu,chao.tian\}@tamu.edu)}
\thanks{C.~Shen is with School of Information Science and Technology, University of Science and Technology of China (USTC), Hefei, China 230027, (e-mail: congshen@ustc.edu.cn).}}

\textfloatsep=0.15cm
\intextsep=0.15cm
\abovecaptionskip=0.15cm
\belowcaptionskip=0.15cm
\setlength{\abovedisplayskip}{5pt}
\setlength{\belowdisplayskip}{5pt}

\maketitle

\begin{abstract}
The problem of multilevel diversity coding with secure regeneration is revisited. Under the assumption that the eavesdropper can access the repair data for all compromised storage nodes, Shao el al. provided a precise characterization of the minimum-bandwidth-regeneration (MBR) point of the achievable normalized storage-capacity repair-bandwidth tradeoff region. In this paper, it is shown that the MBR point of the achievable normalized storage-capacity repair-bandwidth tradeoff region remains the same even if we assume that the eavesdropper can access the repair data for some compromised storage nodes (type \uppercase\expandafter{\romannumeral2} compromised nodes) but only the data contents of the remaining compromised nodes (type \uppercase\expandafter{\romannumeral1} compromised nodes), as long as the number of type \uppercase\expandafter{\romannumeral1} compromised nodes is no greater than that of type \uppercase\expandafter{\romannumeral2} compromised nodes.
\end{abstract}

\section{Introduction}\label{sec:Intro}
Diversity coding, node repair, and security are three basic ingredients of modern distributed storage systems. The interplay of all three ingredients is captured by a fairly general mathematical model known {\em multilevel diversity coding with secure regeneration (MDC-SR)} \cite{Shao-Arxiv17}.

More specifically, in an $(n,d,\ell)$ MDC-SR problem, a total of $d-\ell$ independent files $\mathsf{M}_{\ell+1},\ldots,\mathsf{M}_d$ of size $B_{\ell+1},\ldots,B_d$, respectively, are to be encoded and stored in $n$ distributed storage nodes, each of capacity $\alpha$. The encoding needs to ensure that:
\begin{itemize}
\item (Diversity coding) the file $\mathsf{M}_j$ can be perfectly recovered by having full access to any $j$ out of the total $n$ storage nodes for any $j\in\{\ell+1,\ldots,d\}$;
\vspace{4pt}
\item (Node repair) when node failures occur and there are $d$ remaining nodes in the system, any failed node can be recovered by downloading data of size $\beta$ from each one of the remaining nodes;
\vspace{4pt}
\item (Security) the files $\mathsf{M}_{\ell+1},\ldots,\mathsf{M}_d$ needs to be kept {\em information-theoretically} secure against an eavesdropper, which can access the repair data for $\ell$ compromised storage nodes.
\end{itemize}
Setting $\ell=0$, the above problem reduces to the problem of {\em multilevel diversity coding with regeneration (MDC-R)} considered in \cite{Tian-IT16,Shao-CISS16}. Setting $B_j=0$ for all $j\neq k$, the above problem reduces to the $(n,k,d,\ell)$ {\em secure regenerating code (SRC)} problem considered in \cite{Pawar-ISIT10,Pawar-IT11,Shah-Globecom11,Goparaju-NetCod13,Rawat-IT14,Tandon-IT16,Ye-ISIT16,Shao-IT17}. The goal is to understand the optimal tradeoffs between the storage capacity and repair bandwidth in satisfying all three aforementioned requirements.

From the code construction perspective, it is natural to consider the so-called {\em separate coding} scheme, i.e., to construct a code for the $(n,d,\ell)$ MDC-SR problem, we can simply use an $(n,j,d,\ell)$ SRC to encode the file $\mathsf{M}_j$ for each $j\in\{\ell+1,\ldots,d\}$, and the coded messages for each file remain separate when stored in the storage nodes and during the repair processes. However, despite being a natural scheme, it was shown in \cite{Tian-IT16} that separate coding is in general {\em suboptimal} in achieving the optimal tradeoffs between the normalized storage-capacity and repair-bandwidth for the MDC-R problem (which is a special case of the MDC-SR problem as mentioned previously). On the other hand, it has been shown \cite{Shao-Arxiv17} that separate coding can, in fact, achieve the {\em minimum-bandwidth-regenerating (MBR)} point of the achievable normalized storage-capacity and repair-bandwidth tradeoff region for the general MDC-SR problem. Nevertheless, the optimal tradeoffs between the storage capacity $\alpha$ and download bandwidth $\beta$, and, the performance of the {\em minimum-storage-regenerating (MSR)} point are still not fully understood. Especially for the MSR point, a code was given in \cite{Shah-Globecom11} for SRC problem by extending the known MSR code without any security constraint. This coding scheme can achieve the MSR point when $d\geq 2k-2$ and the eavesdropper can only observe type \uppercase\expandafter{\romannumeral1} compromised nodes (the definition of type \uppercase\expandafter{\romannumeral1} compromised node will be defined in the following part). However, it is still unknown as to whether this code is optimal for the more general eavesdropper model in our paper.

In this paper, we shall revisit the MDC-SR problem with a more general eavesdropping model. More specifically, instead of assuming that the eavesdropper can access the repair data for all compromised storage nodes, we shall assume that the compromised storage nodes can be divided into two different categories: type \uppercase\expandafter{\romannumeral1} compromised nodes and type \uppercase\expandafter{\romannumeral2} compromised nodes. While for the type \uppercase\expandafter{\romannumeral2} compromised nodes, we assume that the eavesdropper can access the repair data as before, for the type \uppercase\expandafter{\romannumeral1} compromised nodes we assume that the eavesdropper can only access the stored data contents.

Let $\ell_1$ and $\ell_2$ be the number of type \uppercase\expandafter{\romannumeral1} compromised nodes and type \uppercase\expandafter{\romannumeral2} compromised nodes respectively, and $\ell:=\ell_1+\ell_2$ be the total number of compromised nodes. By the node repair requirement, the data contents stored at any node can be fully recovered from its repair data. Therefore, for any fixed $\ell$, the eavesdropper becomes weaker as $\ell_1$ increases, which leads to a potentially larger achievable normalized storage-capacity and repair-bandwidth tradeoff region. A question of fundamental interest is to understand whether increasing $\ell_1$ can lead to a {\em strictly} larger achievable normalized storage-capacity and repair-bandwidth tradeoff region. Our main result of the paper is to show that the MBR point of the achievable normalized storage-capacity and repair-bandwidth tradeoff region remains the {\em same}, as long as $\ell_1 \leq \ell/2$ (or equivalently, $\ell_1 \leq \ell_2$ by the fact that $\ell_2=\ell-\ell_1$). From the technical viewpoint, this is mainly accomplished by establishing two outer bounds (one of them must be ``horizontal", i.e., on the normalized repair-bandwidth {\em only}) on the achievable normalized storage-capacity and repair-bandwidth tradeoff region, which intersect precisely at the MBR point.

The rest of the paper is organized as follows. In Section~\ref{sec:PF} we formally introduce the problem of MDC-SR with the generalized eavesdropping model. The main results of the paper are then presented in Section~\ref{sec:Main}. In Section~\ref{sec:Proof}, we introduce two ``exchange" lemmas and use them to establish the main results of the paper. Finally, we conclude the paper in Section~\ref{sec:Con}.

{\em Notation}. Sets and random variables will be written in calligraphic and sans-serif fonts respectively, to differentiate from the real numbers written in normal math fonts. For any two integers $t \leq t'$, we shall denote the set of consecutive integers $\{t,t+1,\ldots,t'\}$ by $[t:t']$. The use of the brackets will be suppressed otherwise.

\section{The Generalized MDC-SR Problem}\label{sec:PF}
In this paper, we study a distributed storage system that share the same file recovery and node repair function with \cite{Shao-Arxiv17}. Let $(n,d,N_1,\ldots,N_d,K,T,S)$ be a tuple of positive integers such that $d<n$. Formally, an $(n,d,N_1,\ldots,N_d,K,T,S)$ code consists of:
\begin{itemize}
\item for each $i \in [1:n]$, a {\em message-encoding} function $f_i: \left(\prod_{j=1}^{d}[1:N_j]\right)\times[1:K] \rightarrow [1:T]$;
\vspace{4pt}
\item for each $\mathcal{A} \subseteq [1:n]: |\mathcal{A}|\in[1:d]$, a {\em message-decoding} function $g_\mathcal{A}: [1:T]^{|\mathcal{A}|} \rightarrow [1:N_{|\mathcal{A}|}]$;
\vspace{4pt}
\item for each $\mathcal{B}\subseteq [1:n]: |\mathcal{B}|=d$, $i' \in \mathcal{B}$, and $i \in [1:n]\setminus\mathcal{B}$, a {\em repair-encoding} function $f^{\mathcal{B}}_{i' \rightarrow i}: [1:T] \rightarrow [1:S]$;
\vspace{4pt}
\item for each $\mathcal{B}\subseteq [1:n]: |\mathcal{B}|=d$ and $i \in [1:n]\setminus\mathcal{B}$, a {\em repair-decoding} function $g^{\mathcal{B}}_i:[1:S]^d \rightarrow [1:T]$.
\end{itemize}

For each $j\in[1:d]$, let $\mathsf{M}_j$ be a message that is uniformly distributed over $[1:N_j]$. The messages $\mathsf{M}_1,\ldots,\mathsf{M}_d$ are assumed to be mutually independent. Let $\mathsf{K}$ be a random key that is uniformly distributed over $[1:K]$ and independent of the messages $(\mathsf{M}_1,\ldots,\mathsf{M}_d)$. For each $i\in[1:n]$, let $\mathsf{W}_i=f_i(\mathsf{M}_1,\ldots,\mathsf{M}_d,\mathsf{K})$ be the data stored at the $i$th storage node, and for each $\mathcal{B}\subseteq [1:n]: |\mathcal{B}|=d$, $i' \in \mathcal{B}$, and $i \in [1:n]\setminus\mathcal{B}$, let $\mathsf{S}^{\mathcal{B}}_{i'\rightarrow i}=f^{\mathcal{B}}_{i'\rightarrow i}(\mathsf{W}_{i'})$ be the data downloaded from the $i'$th storage node in order to regenerate the data originally stored at the $i$th storage node under the context of repair group $\mathcal{B}$. Obviously,
\begin{align*}
(B_j&=\log{N_j}: j\in[1:d]), \quad \alpha=\log{T}, \quad \mbox{and} \;\; \beta=\log{S}
\end{align*}
represent the message sizes, storage capacity, and repair bandwidth, respectively.

The main deference between our definition in this paper and that in \cite{Shao-Arxiv17} is the model of eavesdropper. The eavesdropper now can observer a more complicated data combination consisted of both stored content and repair content.Let $\ell_1$ and $\ell_2$ be two nonnegative integers such that $\ell:=\ell_1+\ell_2<d$. A normalized message-rate storage-capacity repair-bandwidth tuple
 $(\bar{B}_{\ell+1},$ $\ldots,\bar{B}_d,\bar{\alpha},\bar{\beta})$ is said to be {\em achievable} for the $(n,d,\ell_1,\ell_2)$ generalized MDC-SR problem if an
 $(n,d,1,\ldots,1,N_{\ell+1},\ldots,N_d,K,T,S)$ code (i.e., $N_j=1$ for all $j\in[1:\ell]$) can be found such that:
\begin{itemize}
\item (rate normalization)
\begin{align}
\frac{\alpha}{\sum_{t=\ell+1}^dB_t}=\bar{\alpha}, \;
\frac{\beta}{\sum_{t=\ell+1}^dB_t}=\bar{\beta}, \; \frac{B_j}{\sum_{t=\ell+1}^dB_t}=\bar{B}_j\label{eq:Rate}
\end{align}
for any $j\in[\ell+1:d]$;
\vspace{4pt}
\item (message recovery)
\begin{align}
\mathsf{M}_{|\mathcal{A}|} =g_\mathcal{A}(\mathsf{W}_i:i \in\mathcal{A})
\label{eq:MessageRecovery}
\end{align}
for any $\mathcal{A} \subseteq [1:n]: |\mathcal{A}|\in[\ell+1:d]$;
\vspace{4pt}
\item (node regeneration)
\begin{align}
\mathsf{W}_i =g^{\mathcal{B}}_i(\mathsf{S}^{\mathcal{B}}_{i' \rightarrow i}:i'\in \mathcal{B})
\label{eq:NodeRegen}
\end{align}
for any $\mathcal{B}\subseteq [1:n]: |\mathcal{B}|=d$ and $i \in [1:n]\setminus\mathcal{B}$;
\vspace{4pt}
\item (repair secrecy)
\begin{align}
I((\mathsf{M}_{\ell+1},\ldots,\mathsf{M}_d);(\mathsf{W}_i :i\in \mathcal{E}_1), (\mathsf{S}_{\rightarrow j}: j\in \mathcal{E}_2))=0\label{eq:RepairSecrecy}
\end{align}
for any $\mathcal{E}_1,\mathcal{E}_2\subseteq [1:n]$ such that $|\mathcal{E}_1|=\ell_1$,  $|\mathcal{E}_2|=\ell_2$ and $\mathcal{E}_1\cap \mathcal{E}_2=\emptyset$ (so $\mathcal{E}_1$ and $\mathcal{E}_2$ represent the sets of type \uppercase\expandafter{\romannumeral1} and type \uppercase\expandafter{\romannumeral2} compromised storage nodes, respectively), where $\mathsf{S}_{\rightarrow i} :=(\mathsf{S}^{\mathcal{B}}_{i'\rightarrow i}:\mathcal{B}\subseteq [1:n], \; |\mathcal{B}|=d, \; \mathcal{B}\not\ni i, \; i' \in \mathcal{B})$ is the collection of data that can be downloaded from the other nodes to regenerate node $i$.
\end{itemize}
The closure of all achievable $(\bar{B}_{\ell+1},\ldots,\bar{B}_d,\bar{\alpha},\bar{\beta})$ tuples is the {\em achievable normalized message-rate storage-capacity repair-bandwidth tradeoff region} $\mathcal{R}_{n,d,\ell_1,\ell_2}$ for the $(n,d,\ell_1,\ell_2)$ generalized MDC-SR problem. For a fixed normalized message-rate tuple $(\bar{B}_{\ell+1},\ldots,\bar{B}_d)$, the {\em achievable normalized storage-capacity repair-bandwidth tradeoff region} is the collection of all normalized storage-capacity repair-bandwidth pairs $(\bar{\alpha},\bar{\beta})$ such that $(\bar{B}_{\ell+1},$ $\ldots,\bar{B}_d,\bar{\alpha},\bar{\beta}) \in \mathcal{R}_{n,d,\ell_1\ell_2}$ and is denoted by $\mathcal{R}_{n,d,\ell_1,\ell_2}$ $(\bar{B}_{\ell+1},$ $\ldots,\bar{B}_d)$.

Fixing $\ell$ and setting $\ell_1=0$, the $(n,d,\ell_1,\ell_2)$ generalized MDC-SR problem reduces to the $(n,d,\ell)$ MDC-SR problem considered previously in \cite{Shao-Arxiv17}, where it was shown that any achievable normalized message-rate storage-capacity repair-bandwidth tuple $(\bar{B}_{\ell+1},\ldots,\bar{B}_d,\bar{\alpha},\bar{\beta}) \in \mathcal{R}_{n,d,\ell}$ must satisfy:
\begin{align}
\bar{\beta} & \geq \sum_{j=\ell+1}^{d}T_{d,j,\ell}^{-1}\bar{B}_j \label{eq:type2-1}\\
\mbox{and} \quad \bar{\alpha}+(d(d-\ell)-\ell)\bar{\beta} & \geq (d-\ell)(d+1)\sum_{j=\ell+1}^{d}T_{d,j,\ell}^{-1}\bar{B}_j\label{eq:type2-2}.
\end{align}
When set as equalities, the intersection of \eqref{eq:type2-1} and \eqref{eq:type2-2} is given by:
\begin{align}
\label{eq:MBR}
\left(\bar{\alpha},\bar{\beta}\right)
&=\left(d\sum_{j=\ell+1}^{d}T_{d,j,\ell}^{-1}\bar{B}_j,\sum_{j=\ell+1}^{d}T_{d,j,\ell}^{-1}\bar{B}_j\right)
\end{align}
which can be achieved by separate encoding with a previous scheme proposed by Shah, Rashmi, and Kumar \cite{Shah-Globecom11}. This provides a precise characterization of the MBR point for the $(n,d,\ell)$ MDC-SR problem.

\section{Main Results}\label{sec:Main}
Our main result of the paper is to show that the tradeoff point \eqref{eq:MBR} remains to be the MBR point of $\mathcal{R}_{n,d,\ell_1,\ell_2}$ for the generalized MDC-SR problem as long as $\ell_1 \leq \ell_2$. The results are summarized in the following theorem.

\begin{thm}\label{thm}
For the generalized MDC-SR problem, any achievable normalized message-rate storage-capacity repair-bandwidth tuple $(\bar{B}_{\ell+1},\ldots,\bar{B}_d,\bar{\alpha},\bar{\beta}) \in $
$\mathcal{R}_{n,d,\ell_1,\ell_2}$ must satisfy:
\begin{align}
\bar{\beta}  \geq \sum_{j=\ell+1}^{d}T_{d,j,\ell}^{-1}\bar{B}_j \label{eq:B3}
\end{align}
and in addition, when $\ell_1 \leq \ell_2=\ell-\ell_1$, we also have
\begin{align}
\bar{\alpha}+T_{d,d,\ell_1+1}\bar{\beta} & \geq (T_{d,d,\ell_1}+\ell_1)\sum_{j=\ell+1}^{d}T_{d,j,\ell}^{-1}\bar{B}_j\label{eq:B4}
\end{align}
where $T_{d,k,\ell}:=\sum_{t=\ell+1}^{k}(d+1-t)$. When set as equalities, the intersection of \eqref{eq:B3} and \eqref{eq:B4} is precisely given by \eqref{eq:MBR}. We may thus conclude immediately that \eqref{eq:MBR} is the MBR point of $\mathcal{R}_{n,d,\ell_1,\ell_2}$ for the generalized MDC-SR problem as long as $\ell_1 \leq \ell_2$.
\end{thm}

\begin{figure}[htbp]
   \centering
   \includegraphics[width=0.5\textwidth]{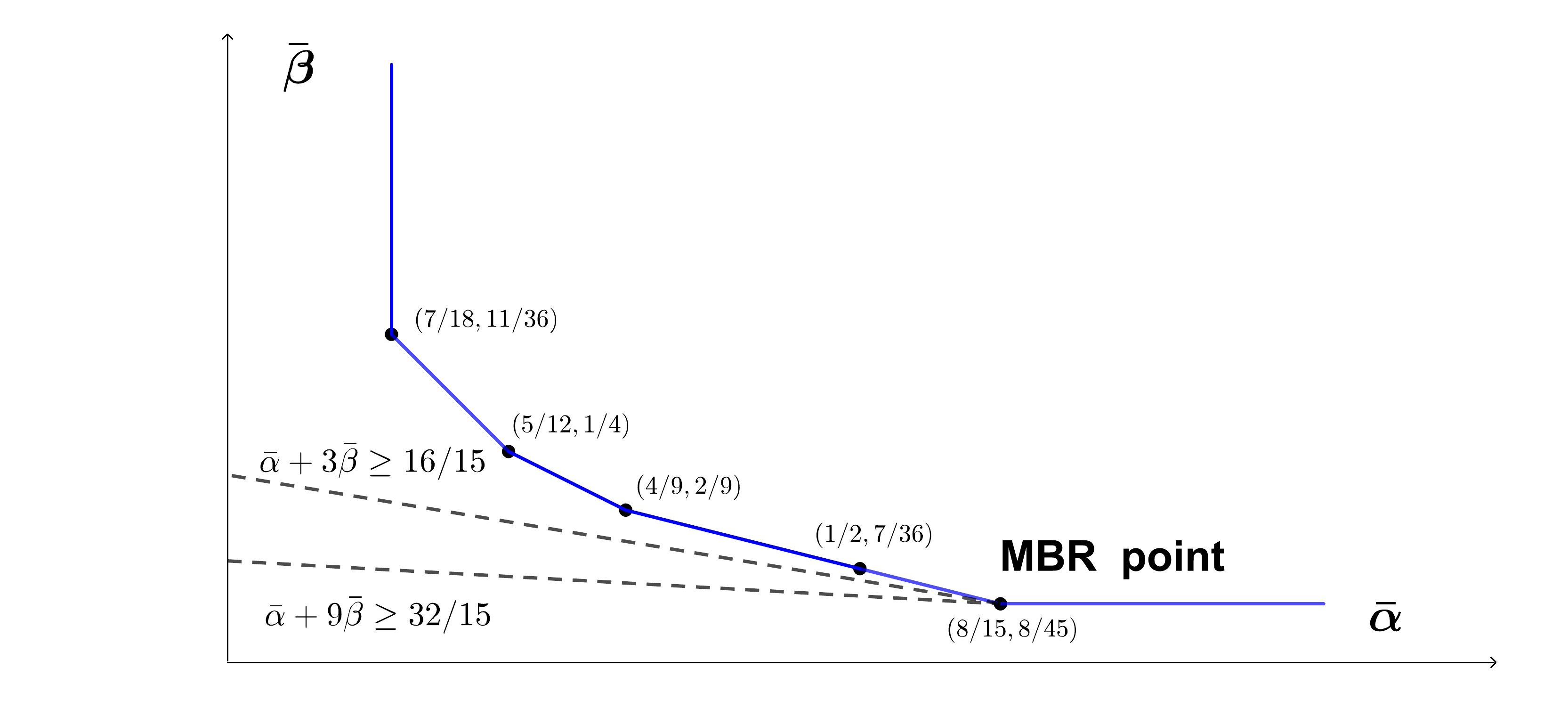}
  \caption{The optimal tradeoff region for $(4,3,0,0)$ MDC-SR problem when $(\bar{B}_1,\bar{B}_2,\bar{B}_3)=(0,1/3,2/3)$ . The outer bounds \eqref{eq:B3}, \eqref{eq:B} and \eqref{eq:type2-2} are evaluated as $\bar{\beta} \geq 8/45$, $\bar{\alpha}+3\bar{\beta} \geq 16/15$, and $\bar{\alpha}+9\bar{\beta} \geq 32/15$, respectively. When set as equalities, they intersect precisely at the MBR point $(8/15,8/45)$.}
   \label{fig1}
 \end{figure}

Note that setting $\ell_1=0$, the outer bound \eqref{eq:B4} reduces to
\begin{align}
\bar{\alpha}+\frac{d(d-1)}{2}\bar{\beta} & \geq \frac{d(d+1)}{2}\sum_{j=\ell+1}^{d}T_{d,j,\ell}^{-1}\bar{B}_j.
\label{eq:B}
\end{align}
So while the outer bound \eqref{eq:B3} coincides with \eqref{eq:type2-1}, the outer bound \eqref{eq:B4} does {\em not} reduce to \eqref{eq:type2-2} when setting $\ell_1=0$. Simple calculations yield that the outer bound \eqref{eq:B} is stronger than \eqref{eq:type2-2} if and only if $\ell\leq d/2$. In particular, when $\ell=0$, the outer bound \eqref{eq:B} reduces to that for the $(n,d)$ MDC-R problem \cite{Shao-CISS16}, while the outer bound \eqref{eq:type2-2} is {\em strictly} weaker. Fig. \ref{fig1} shows the comparison of \eqref{eq:B} and \eqref{eq:type2-2} when $(\bar{B}_1,\bar{B}_2,\bar{B}_3)=(0,1/3,2/3)$ in $(4,3,0,0)$ MDC-SR problem. In this figure, the outer bound \eqref{eq:type2-2} is below outer bound \eqref{eq:B}, though both of them intersect with \eqref{eq:B3} at the MBR point.

\section{Proof of the Main Results}\label{sec:Proof}
Let us first outline the main ingredients for proving the outer bounds \eqref{eq:B3} and \eqref{eq:B4}.

\begin{itemize}
\item[1)] {\em Total number of nodes.} To prove the outer bounds \eqref{eq:B3} and \eqref{eq:B4}, let us first note that these bounds are {\em independent} of the total number of storage nodes $n$ in the system. Therefore, in our proof, we only need to consider the cases where $n=d+1$. For the cases where $n>d+1$, since any subsystem consisting of $d+1$ out of the total $n$ storage nodes must give rise to a $(d+1,d,\ell)$ MDC-SR problem. Therefore, these {\em outer} bounds must apply as well. When $n=d+1$, any repair group $\mathcal{B}$ of size $d$ is uniquely determined by the node $j$ to be repaired, i.e., $\mathcal{B}=[1:n]\setminus\{j\}$, and hence can be dropped from the notation $\mathsf{S}^{\mathcal{B}}_{i \rightarrow j}$ without causing any confusion.
\item[2)] {\em Code symmetry.} Due to the built-in {\em symmetry} of the problem, to prove the outer bounds \eqref{eq:B3} and \eqref{eq:B4}, we only need to consider the so-called {\em symmetrical} codes \cite{Tian-JSAC13} for which the joint entropy of any subset of random variables from
\begin{align*}
&\left((\mathsf{M}_1,\ldots,\mathsf{M}_d),\mathsf{K},\right.\\
&\left.\hspace{30pt}(\mathsf{W}_i:i\in[1:n]),(\mathsf{S}_{i \rightarrow j}: i,j\in[1:n],i\neq j)\right)
\end{align*}
remains {\em unchanged} under {\em any} permutation over the {\em storage-node} indices.
\item[3)] {\em Key collections of random variables.} Focusing on the symmetrical $(n=d+1,d,N_1,\ldots,N_d,K,T,S)$ codes, the following collections of random variables play a key role in our proof:
\begin{align*}
& \mathsf{M}_{\mathcal{A}} := (\mathsf{M}_i: i \in \mathcal{A}), \quad \mathcal{A}\subseteq [1:d]\\
& \mathsf{M}^{(m)} := \mathsf{M}_{[1:m]}, \quad m\in[1:d]\\
&\mathsf{W}_{\mathcal{A}} :=\left(\mathsf{W}_i:i\in \mathcal{A}\right), \quad \mathcal{A}\subseteq [1:n]\\
&\mathsf{S}_{i\rightarrow\mathcal{B}} := \left(\mathsf{S}_{i \rightarrow j}: j\in\mathcal{B}\right),\quad i\in [1:n], \; \mathcal{B}\subseteq [1:n]\setminus \{i\}\\
&\mathsf{S}_{\mathcal{B}\rightarrow j} := \left(\mathsf{S}_{i \rightarrow j}: \quad i\in\mathcal{B}\right), j\in[1:n],\; \mathcal{B}\subseteq [1:n]\setminus \{j\}\\
&\mathsf{S}_{\rightarrow j} := \mathsf{S}_{[1:j-1]\cup[j+1:n]\rightarrow j}, \quad j\in[1:n]\\
&\mathsf{S}_{\rightarrow \mathcal{B}} :=\left(\mathsf{S}_{\rightarrow j}:j\in\mathcal{B}\right), \quad \mathcal{B}\subseteq [1:n]\\
&\underline{\mathsf{S}}_{\rightarrow j} := \mathsf{S}_{[1:j-1]\rightarrow j}, \quad j\in[1:n]\\
&\underline{\mathsf{S}}_{\rightarrow \mathcal{B}} :=(\underline{\mathsf{S}}_{\rightarrow j}:j\in\mathcal{B}), \quad \mathcal{B}\subseteq [1:n]\\
&\overline{\mathsf{S}}_{\rightarrow j} := \mathsf{S}_{[j+1:n]\rightarrow j}, \quad j\in[1:n]\\
&\overline{\mathsf{S}}_{\rightarrow \mathcal{B}} :=(\overline{\mathsf{S}}_{\rightarrow j}:j\in\mathcal{B}), \quad \mathcal{B}\subseteq [1:n]\\
&\mathsf{U}^{(t,s)} :=(\mathsf{W}_{[1:t]},\overline{\mathsf{S}}_{\rightarrow[t+1:s]}), \quad s\in[1:n], \; t\in[0:s]\\
&\mathsf{U}^{(s)} := \mathsf{U}^{(0,s)}.
\end{align*}
These collections of random variables have also been used in \cite{Shao-IT17,Shao-CISS16}.
\end{itemize}

An important part of the proof is to understand the relations between the collections of random variables defined above, and to use them to derive the desired converse results. We shall discuss this next.

\subsection{Technical Lemmas}
\begin{lemma}\label{lemma1}
For any $(n=d+1,d,N_1,\ldots,N_d,K,T,S)$ code that satisfies the node regeneration requirement \eqref{eq:NodeRegen}, $(\underline{\mathsf{S}}_{\rightarrow[t+1:s]},\mathsf{W}_{[t+1:s]})$ is a function of $\mathsf{U}^{(t,s)}$ for any $s \in [1:n]$ and $t\in[0:s-1]$.
\end{lemma}


The above lemma, which was first introduced in \cite{Shao-Arxiv17,Shao-IT17}, demonstrates the ``compactness" of $\mathsf{U}^{(t,s)}$ and has a number of direct consequences. For example, for any fixed $s \in [1:n]$, it is clear from Lemma~\ref{lemma1} that $\mathsf{U}^{(t_2,s)}$ is a function of $\mathsf{U}^{(t_1,s)}$ and hence $H(\mathsf{U}^{(t_2,s)}) \leq H(\mathsf{U}^{(t_1,s)})$ for any $0 \leq t_1 \leq t_2 \leq s-1$.

\begin{lemma}[Exchange lemma \uppercase\expandafter{\romannumeral1} \cite{Shao-Arxiv17}]
\label{lemma:exchange}
For any symmetrical $(n=d+1,d,N_1,\ldots,N_d$, $K,T,S)$ code that satisfies the node regeneration requirement \eqref{eq:NodeRegen}, we have
\begin{align}
& \frac{d+1-j}{d-m}H(\mathsf{U}^{(i,m)}|\mathsf{M}^{(m)})+H(\mathsf{U}^{(i',j)}|\mathsf{M}^{(m)})\nonumber\\
& \hspace{15pt} \geq \frac{d+1-j}{d-m}H(\mathsf{U}^{(i,m+1)}|\mathsf{M}^{(m)})+H(\mathsf{U}^{(i',j-1)}|\mathsf{M}^{(m)})\label{eq:exchange}
\end{align}
for any $m\in[1:d-1]$, $i\in[0:m-1]$, $i'\in[0:i]$, and $j\in[i'+1:m-i+i'+1]$.
\end{lemma}
\begin{coro}
\label{coro1}
For any symmetrical $(n=d+1,d,N_1,\ldots, N_d$, $K,T,S)$ code that satisfies the node regeneration requirement \eqref{eq:NodeRegen}, we have
\begin{align}
& \frac{T_{d,j_1,j_2}}{d-j_1}H(\mathsf{U}^{(i,j_1)}|\mathsf{M}^{(j_1)})+H(\mathsf{U}^{(i',j_1)}|\mathsf{M}^{(j_1)})\nonumber\\
& \hspace{15pt} \geq \frac{T_{d,j_1,j_2}}{d-j_1}H(\mathsf{U}^{(i,j_1+1)}|\mathsf{M}^{(j_1)})+H(\mathsf{U}^{(i',j_2)}|\mathsf{M}^{(j_1)})\label{eq:coro1}
\end{align}
for any $j_1\in[1:d-1]$, $i=[0:j_1]$, $i' \in[\max\{0,i-1\}:i]$ and $j_2\in[i':j_1-1]$.
\end{coro}

\begin{IEEEproof}
Set $m=j_1$ in \eqref{eq:exchange}. We have
\begin{align}
& \frac{d+1-j}{d-j_1}H(\mathsf{U}^{(i,j_1)}|\mathsf{M}^{(j_1)})+H(\mathsf{U}^{(i',j)}|\mathsf{M}^{(j_1)})\nonumber\\
& \hspace{10pt}\geq \frac{d+1-j}{d-j_1}H(\mathsf{U}^{(i,j_1+1)}|\mathsf{M}^{(j_1)})+H(\mathsf{U}^{(i',j-1)}|\mathsf{M}^{(j_1)})\label{eq:NewExchange2}
\end{align}
for any $j\in[j_2+1:j_1]$. Add the inequalities \eqref{eq:NewExchange2} for $j\in[j_2+1:j_1]$ and cancel the common term $\sum_{j=j_2+1}^{j_1-1}H(\mathsf{U}^{(i',j)}|\mathsf{M}^{(j_1)})$ from both sides. We have
\begin{align}
&\frac{T_{d,j_1,j_2}}{d-j_1}H(\mathsf{U}^{(i,j_1)}|\mathsf{M}^{(j_1)})+H(\mathsf{U}^{(i',j_1)}|\mathsf{M}^{(j_1)})\nonumber\\
& \hspace{15pt} \geq \frac{T_{d,j_1,j_2}}{d-j_1}H(\mathsf{U}^{(i,j_1+1)}|\mathsf{M}^{(j_1)})+H(\mathsf{U}^{(i',j_2)}|\mathsf{M}^{(j_1)}).\nonumber
\end{align}
\end{IEEEproof}

\begin{coro}
\label{coro2}
For any symmetrical $(n=d+1,d,N_1,\ldots,N_d$, $K,T,S)$ code that satisfies the node regeneration requirement \eqref{eq:NodeRegen}, we have
\begin{align}
T_{d,m,\ell}^{-1}H(\mathsf{U}^{(\ell_1,m)}|\mathsf{M}^{(m)})&\geq T_{d,m+1,\ell}^{-1}H(\mathsf{U}^{(\ell_1,m+1)}|\mathsf{M}^{(m)})+\nonumber\\
 &
 (T_{d,m,\ell}^{-1}-T_{d,m+1,\ell}^{-1})H(\mathsf{U}^{(\ell_1,\ell)}|\mathsf{M}^{(m)})\label{eq:coro2}
\end{align}
for any $\ell\in[0:d-1]$, $\ell_1\in[0:\ell]$ and $m\in[\ell+1:d-1]$.
\end{coro}

\begin{IEEEproof}
Set $i=i'=\ell_1$, $j_1=m$ and $j_2=\ell$ in \eqref{eq:coro1}. We have
\begin{align}
& \frac{T_{d,m,\ell}}{d-m}H(\mathsf{U}^{(\ell_1,m)}|\mathsf{M}^{(m)})+H(\mathsf{U}^{(\ell_1,m)}|\mathsf{M}^{(m)})\nonumber\\
& \hspace{15pt} \geq \frac{T_{d,m,\ell}}{d-m}H(\mathsf{U}^{(\ell_1,m+1)}|\mathsf{M}^{(m)})+H(\mathsf{U}^{(\ell_1,\ell)}|\mathsf{M}^{(m)}),
\end{align}
which can be equivalently written as
\begin{align}
&\frac{T_{d,m+1,\ell}}{d-m}H(\mathsf{U}^{(\ell_1,m)}|\mathsf{M}^{(m)})\nonumber\\
& \hspace{30pt} \geq \frac{T_{d,m,\ell}}{d-m}H(\mathsf{U}^{(\ell_1,m+1)}|\mathsf{M}^{(m)})+ H(\mathsf{U}^{(\ell_1,\ell)}|\mathsf{M}^{(m)})\label{eq:Temp4}
\end{align}
by the fact that $T_{d,m,\ell}+(d-m)=T_{d,m+1,\ell}$. Multiplying both sides of \eqref{eq:Temp4} by
$$\frac{d-m}{T_{d,m+1,\ell}T_{d,m,\ell}}=T_{d,m,\ell}^{-1}-T_{d,m+1,\ell}^{-1}$$
completes the proof of \eqref{eq:coro1}.
\end{IEEEproof}
\begin{lemma}[exchange lemma \uppercase\expandafter{\romannumeral2}]
\label{lemma:mutant exchange}
For any symmetrical $(n=d+1,d,N_1$, $\ldots,N_d,K,T,S)$ code that satisfies the node regeneration requirement \eqref{eq:NodeRegen}, we have
\begin{align}
& \frac{d-\ell_1}{d-\ell}H(\mathsf{U}^{(\ell_1,\ell)})+H(\mathsf{U}^{(\ell_1,\ell_1+1)},\mathsf{S}_{\ell_1+1\rightarrow [1:\ell_1]})\nonumber\\
& \hspace{15pt} \geq \frac{d-\ell_1}{d-\ell}H(\mathsf{U}^{(\ell_1,\ell+1)})+H(\mathsf{U}^{(\ell_1,\ell_1)},\mathsf{S}_{\ell_1+1\rightarrow [1:\ell_1]})\label{eq:mutant exchange}
\end{align}
for any $\ell \in[1:d-1]$ and $\ell_1 \in[0:\left\lfloor \ell/2 \right\rfloor]$.
\end{lemma}

\begin{IEEEproof}
See the Appendix.
\end{IEEEproof}

We note here that when setting $\ell_1=0$, the above lemma coincides with Lemma~\ref{lemma:exchange} with $i=i'=0$ and $j=1$.

\subsection{The Proof}
Consider a symmetrical $(n=d+1,d,1,\ldots,1,N_{\ell+1},\ldots,N_d$, $K,T,S)$ regenerating code that satisfies the rate normalization requirement \eqref{eq:Rate}, the message recovery requirement \eqref{eq:MessageRecovery}, the node regeneration requirement \eqref{eq:NodeRegen}, and the repair secrecy requirement \eqref{eq:RepairSecrecy}. Let us first prove a few intermediate results. The outer bounds \eqref{eq:B3} and \eqref{eq:B4} will then follow immediately.

\begin{prop}\label{prop1}
\begin{align}
&\frac{1}{d-\ell}H(\mathsf{U}^{(\ell_1,\ell+1)}) \geq \sum_{j=\ell+1}^{m}T_{d,j,\ell}^{-1}B_j+\nonumber\\
& \hspace{15pt} T_{d,m,\ell}^{-1}H(\mathsf{U}^{(\ell_1,m)}|\mathsf{M}_{[\ell+1:m]})+\left(\frac{1}{d-\ell}-T_{d,m,\ell}^{-1}\right)H(\mathsf{U}^{(\ell_1,\ell)})\label{eq:EG}
\end{align}
for any $m\in [\ell+1:d]$. Consequently,
\begin{align}
\frac{1}{d-\ell}H(\mathsf{U}^{(\ell_1,\ell+1)}) \geq \sum_{j=\ell+1}^{d}T_{d,j,\ell}^{-1}B_j+\frac{1}{d-\ell}H(\mathsf{U}^{(\ell_1,\ell)}).
\label{eq:prop1}
\end{align}
\end{prop}

\begin{IEEEproof}
To see \eqref{eq:EG}, consider proof by induction. For the base case with $m=\ell+1$, we have
\begin{align*}
&\frac{1}{d-\ell}H(\mathsf{U}^{(\ell_1,\ell+1)})\nonumber\\
 \stackrel{(a)}{=}& \frac{1}{d-\ell}H(\mathsf{U}^{(\ell_1,\ell+1)},\mathsf{M}_{\ell+1})\\
 \stackrel{(b)}{=}& \frac{1}{d-\ell}\left(H(\mathsf{M}_{\ell+1})+H(\mathsf{U}^{(\ell_1,\ell+1)}|\mathsf{M}_{\ell+1})\right)\\
 \stackrel{(c)}{=}& \frac{1}{d-\ell}\left(B_{\ell+1}+H(\mathsf{U}^{(\ell_1,\ell+1)}|\mathsf{M}_{\ell+1})\right)\\
 \stackrel{(d)}{=}& T_{d,\ell+1,\ell}^{-1}B_{\ell+1}+T_{d,\ell+1,\ell}^{-1}H(\mathsf{U}^{(\ell_1,\ell+1)}|\mathsf{M}_{\ell+1})
\end{align*}
where $(a)$ follows from the fact that $\mathsf{M}_{\ell+1}$ is a function of $\mathsf{W}_{[1:\ell+1]}$, which is a function of $\mathsf{U}^{(\ell_1,\ell+1)}$ by Lemma~\ref{lemma1}; $(b)$ follows from the chain rule for entropy; $(c)$ follows from the fact that $H(\mathsf{M}_{\ell+1})=B_{\ell+1}$; and $(d)$ follows from the fact that $T_{d,\ell+1,\ell}=d-\ell$. Assuming that \eqref{eq:EG} holds for some $m\in [\ell+1:d-1]$, we have
\begin{align*}
& \frac{1}{d-\ell}H(\mathsf{U}^{(\ell_1,\ell+1)})\\
& \stackrel{(a)}{\geq} \sum_{j=\ell+1}^{m}T_{d,j,\ell}^{-1}B_j+T_{d,m,\ell}^{-1}H(\mathsf{U}^{(\ell_1,m)}|\mathsf{M}_{[\ell+1:m]})+\\
& \hspace{20pt} \left(\frac{1}{d-\ell}-T_{d,m,\ell}^{-1}\right)H(\mathsf{U}^{(\ell_1,\ell)})\\
& \stackrel{(b)}{\geq} \sum_{j=\ell+1}^{m}T_{d,j,\ell}^{-1}B_j+T_{d,m+1,\ell}^{-1}H(\mathsf{U}^{(\ell_1,m+1)}|\mathsf{M}_{[\ell+1:m]})+\\
& \hspace{20pt} \left(\frac{1}{d-\ell}-T_{d,m+1,\ell}^{-1}\right)H(\mathsf{U}^{(\ell_1,\ell)})\\
& \stackrel{(c)}{\geq} \sum_{j=\ell+1}^{m}T_{d,j,\ell}^{-1}B_j+\nonumber\\
& \hspace{20pt} T_{d,m+1,\ell}^{-1}H(\mathsf{U}^{(\ell_1,m+1)},\mathsf{M}_{m+1}|\mathsf{M}_{[\ell+1:m]})+\\
& \hspace{20pt} \left(\frac{1}{d-\ell}-T_{d,m+1,\ell}^{-1}\right)H(\mathsf{U}^{(\ell_1,\ell)})\\
& \stackrel{(d)}{=} \sum_{j=\ell+1}^{m}T_{d,j,\ell}^{-1}B_j+T_{d,m+1,\ell}^{-1}H(\mathsf{M}_{m+1}|\mathsf{M}_{[\ell+1:m]})+\\
& \hspace{20pt} T_{d,m+1,\ell}^{-1}H(\mathsf{U}^{(\ell_1,m+1)}|\mathsf{M}_{[\ell+1:m+1]})+\\
& \hspace{20pt} \left(\frac{1}{d-\ell}-T_{d,m+1,\ell}^{-1}\right)H(\mathsf{U}^{(\ell_1,\ell)})\\
& \stackrel{(e)}{=} \sum_{j=\ell+1}^{m}T_{d,j,\ell}^{-1}B_j+T_{d,m+1,\ell}^{-1}B_{m+1}+\\
& \hspace{20pt} T_{d,m+1,\ell}^{-1}H(\mathsf{U}^{(\ell_1,m+1)}|\mathsf{M}_{[\ell+1:m+1]})+\\
& \hspace{20pt} \left(\frac{1}{d-\ell}-T_{d,m+1,\ell}^{-1}\right)H(\mathsf{U}^{(\ell_1,\ell)})\\
& = \sum_{j=\ell+1}^{m+1}T_{d,j,\ell}^{-1}+T_{d,m+1,\ell}^{-1}H(\mathsf{U}^{(\ell_1,m+1)}|\mathsf{M}_{[\ell+1:m+1]})+\\
& \hspace{20pt} \left(\frac{1}{d-\ell}-T_{d,m+1,\ell}^{-1}\right)H(\mathsf{U}^{(\ell_1,\ell)})
\end{align*}
where $(a)$ follows from the induction assumption; $(b)$ follows from Corollary~\ref{coro1}; $(c)$ follows from the fact that $\mathsf{M}_{m+1}$ is a function of $\mathsf{W}_{[1:m+1]}$, which is is a function of $\mathsf{U}^{(\ell_1,m+1)}$ by Lemma~\ref{lemma1}; $(d)$ follows from the chain rule for entropy; and $(e)$ follows from the facts that $\mathsf{M}_{m+1}$ is independent of $\mathsf{M}_{[\ell+1:m]}$ and that $H(\mathsf{M}_{m+1})=B_{m+1}$. This completes the induction step and hence the proof of \eqref{eq:EG}.

To see \eqref{eq:prop1}, simply set $m=d$ in \eqref{eq:EG}. We have
\begin{align}
&\frac{1}{d-\ell}H(\mathsf{U}^{(\ell_1,\ell+1)}) \geq \sum_{j=\ell+1}^{d}T_{d,j,\ell}^{-1}B_j+\nonumber\\
& \hspace{20pt} T_{d,d,\ell}^{-1}H(\mathsf{U}^{(\ell_1,d)}|\mathsf{M}_{[\ell+1:d]})+\left(\frac{1}{d-\ell}-T_{d,d,\ell}^{-1}\right)H(\mathsf{U}^{(\ell_1,\ell)}).\label{eq:EG2}
\end{align}
Note that
\begin{align}
H(\mathsf{U}^{(\ell_1,d)}|\mathsf{M}_{[\ell+1:d]}) \geq H(\mathsf{U}^{(\ell_1,\ell)}|\mathsf{M}_{[\ell+1:d]}) = H(\mathsf{U}^{(\ell_1,\ell)})\label{eq:EG3}
\end{align}
where the last equality follows from the fact that $I(\mathsf{U}^{(\ell_1,\ell)};\mathsf{M}_{[\ell+1:d]})=0$ by the repair secrecy requirement \eqref{eq:RepairSecrecy}. Substituting \eqref{eq:EG3} into \eqref{eq:EG2} completes the proof of \eqref{eq:prop1}.
\end{IEEEproof}

\begin{prop}\label{prop2}
\begin{align}
H(\mathsf{S}_{\ell_1+1\rightarrow[1:\ell_1]})+\frac{\ell_1}{d-\ell}H(\mathsf{U}^{(\ell_1,\ell)})\geq
\frac{\ell_1}{d-\ell}H(\mathsf{U}^{(\ell_1,\ell+1)}).\label{eq:prop2}
\end{align}
\end{prop}

\begin{IEEEproof}
First note that for any $m \in [1:\ell_2+1]$ and $k\in[\ell+1:d+1]$, we have
\begin{align}
H&(\mathsf{S}_{\ell_1+1\rightarrow[1:m]})+H(\mathsf{U}^{(\ell_1,\ell)},\mathsf{S}_{[\ell+2:k]\rightarrow \ell+1})\nonumber\\
& \stackrel{(a)}{=} H(\mathsf{S}_{k+1\rightarrow[\ell_1+1:\ell_1+m-1]\cup\{\ell+1\}})+\nonumber\\
& \hspace{20pt} H(\mathsf{U}^{(\ell_1,\ell)},\mathsf{S}_{\rightarrow [\ell_1+1:\ell]},\mathsf{S}_{[\ell+2:k]\rightarrow \ell+1})\nonumber\\
& \stackrel{(b)}{\geq} H(\mathsf{S}_{k+1\rightarrow[\ell_1+1:\ell_1+m-1]})+H(\mathsf{U}^{(\ell_1,\ell)},\mathsf{S}_{[\ell+2:k+1]\rightarrow \ell+1})\nonumber\\
& \stackrel{(c)}{=} H(\mathsf{S}_{\ell_1+1\rightarrow[1:m-1]})+H(\mathsf{U}^{(\ell_1,\ell)},\mathsf{S}_{[\ell+2:k+1]\rightarrow \ell+1})\label{eq:STE}
\end{align}
where $(a)$ and $(c)$ follow from the fact that $H(\mathsf{S}_{\ell_1+1\rightarrow[1:m]})=H(\mathsf{S}_{k+1\rightarrow[1:m-1]\cup\{\ell+1\}})$ and $H(\mathsf{S}_{k+1\rightarrow[1:m-1]})=H(\mathsf{S}_{\ell_1+1\rightarrow[1:m-1]})$ due to the symmetrical code that we consider, and $(b)$ follows from the submodularity of the entropy function. Add \eqref{eq:STE} over $m\in[1:\ell_1]$ and cancel $\sum_{m=1}^{\ell_1-1}H(\mathsf{S}_{d+1\rightarrow[1:m]})$ from both sides. We have
\begin{align}
&H(\mathsf{S}_{d+1\rightarrow[1:\ell]})+\ell_1 H(\mathsf{U}^{(\ell_1,\ell)},\mathsf{S}_{[\ell+2:k+1]\rightarrow \ell+1}) \geq\nonumber\\
 &\quad \ell_1 H(\mathsf{U}^{(\ell_1,\ell)},\mathsf{S}_{[\ell+2:k+1]\rightarrow \ell+1}).\label{eq:STE2}
\end{align}
Add \eqref{eq:STE2} over $k\in [\ell+1:d]$ and cancel $\sum_{k=\ell+1}^{d-1} H(\mathsf{U}^{(\ell_1,\ell)},\mathsf{S}_{[\ell+2:k+1]\rightarrow \ell+1})$ from both sides. We have
\begin{align}
&(d-\ell)H(\mathsf{S}_{d+1\rightarrow[1:\ell]})+\ell_1 H(\mathsf{U}^{(\ell_1,\ell)})\nonumber\\
 \geq& \ell_1 H(\mathsf{U}^{(\ell_1,\ell)},\mathsf{S}_{[\ell+2:d+1]\rightarrow \ell+1})\nonumber\\
  = &\ell_1 H(\mathsf{U}^{(\ell_1,\ell+1)}).\label{eq:STE2}
\end{align}
Multiplying both sides by $ (d-\ell)^{-1}$ completes the proof of \eqref{eq:prop2}

\end{IEEEproof}

\begin{prop}\label{prop3}
\begin{align}
&H(\mathsf{U}^{(\ell_1+1,m)})+\frac{d-m}{d-\ell}H(\mathsf{U}^{(\ell_1,\ell+1)})\nonumber\\
& \geq (d-m)\sum_{j=\ell+1}^{m}T_{d,j,\ell}^{-1}B_j+H(\mathsf{U}^{(\ell_1+1,m+1)})+\frac{d-m}{d-\ell}H(\mathsf{U}^{(\ell_1,\ell)})\label{eq:JH}
\end{align}
for any $m\in[\ell+1:d-1]$. Consequently,
\begin{align}
& H(\mathsf{U}^{(\ell_1+1,\ell+1)})+\frac{T_{d,d,\ell+1}}{d-\ell}H(\mathsf{U}^{(\ell_1,\ell+1)})\nonumber\\
& \hspace{20pt} \geq T_{d,d,\ell}\sum_{j=\ell+1}^{d}T_{d,j,\ell}^{-1}B_j+\frac{T_{d,d,\ell}}{d-\ell}H(\mathsf{U}^{(\ell_1,\ell)}).\label{eq:prop3}
\end{align}
\end{prop}

\begin{IEEEproof}
To see \eqref{eq:JH}, note that for any $m\in[\ell+1:d-1]$, we have
\begin{align*}
&H(\mathsf{U}^{(\ell_1+1,m)}|\mathsf{M}_{[\ell+1:m]})+\frac{d-m}{d-\ell}H(\mathsf{U}^{(\ell_1,\ell+1)})\\
& \stackrel{(a)}{\geq} H(\mathsf{U}^{(\ell_1+1,m)}|\mathsf{M}_{[\ell+1:m]})+ \\
& \hspace{20pt} (d-m)\left(\sum_{j=\ell+1}^{m}T_{d,j,\ell}^{-1}B_j+\right.T_{d,m,\ell}^{-1}H(\mathsf{U}^{(\ell_1,m)}|\mathsf{M}_{[\ell+1:m]})\\
& \hspace{20pt} \left.+\left(\frac{1}{d-\ell}-T_{d,m,\ell}^{-1}\right)H(\mathsf{U}^{(\ell_1,\ell)})\right)\\
& = H(\mathsf{U}^{(\ell_1+1,m)}|\mathsf{M}_{[\ell+1:m]})+\\
& \hspace{20pt}
 (d-m)T_{d,m,\ell}^{-1}H(\mathsf{U}^{(\ell_1,m)}|\mathsf{M}_{[\ell+1:m]})+\\
& \hspace{20pt} (d-m)\left(\sum_{j=\ell+1}^{m}T_{d,j,\ell}^{-1}B_j+\left(\frac{1}{d-\ell}-T_{d,m,\ell}^{-1}\right)H(\mathsf{U}^{(\ell_1,\ell)})\right)\\
& \stackrel{(b)}{\geq} H(\mathsf{U}^{(\ell_1+1,m+1)}|\mathsf{M}_{[\ell+1:m]})+\\
& \hspace{20pt}
(d-m)T_{d,m,\ell}^{-1}H(\mathsf{U}^{(\ell_1,\ell)}|\mathsf{M}_{[\ell+1:m]})+\\
& \hspace{20pt} (d-m)\left(\sum_{j=\ell+1}^{m}T_{d,j,\ell}^{-1}B_j+\left(\frac{1}{d-\ell}-T_{d,m,\ell}^{-1}\right)H(\mathsf{U}^{(\ell_1,\ell)})\right)\\
& \stackrel{(c)}{=} H(\mathsf{U}^{(\ell_1+1,m+1)}|\mathsf{M}_{[\ell+1:m]})+(d-m)T_{d,m,\ell}^{-1}H(\mathsf{U}^{(\ell_1,\ell)})+\\
& \hspace{20pt} (d-m)\left(\sum_{j=\ell+1}^{m}T_{d,j,\ell}^{-1}B_j+\left(\frac{1}{d-\ell}-T_{d,m,\ell}^{-1}\right)H(\mathsf{U}^{(\ell_1\ell)})\right)\\
& = H(\mathsf{U}^{(\ell_1+1,m+1)}|\mathsf{M}_{[\ell+1:m]})+(d-m)\sum_{j=\ell+1}^{m}T_{d,j,\ell}^{-1}B_j+\\
& \hspace{20pt} \frac{d-m}{d-\ell}H(\mathsf{U}^{(\ell_1,\ell)})
\end{align*}
where $(a)$ follows from \eqref{eq:EG} of Proposition~\ref{prop1}; $(b)$ follows from Corollary~\ref{coro2}; and $(c)$ follows from the fact that $I(\mathsf{U}^{(\ell_1,\ell)};\mathsf{M}_{[\ell+1:m]})=0$ due to the repair secrecy requirement \eqref{eq:RepairSecrecy}. Adding $H(\mathsf{M}_{[\ell+1:m]})$ to both sides and using the facts that
\begin{align*}
&H(\mathsf{U}^{(\ell_1+1,m)}|\mathsf{M}_{[\ell+1:m]})+H(\mathsf{M}_{[\ell+1:m]})\\
&\hspace{40pt}=H(\mathsf{U}^{(\ell_1+1,m)},\mathsf{M}_{[\ell+1:m]})\stackrel{(a)}{=}H(\mathsf{U}^{(\ell_1+1,m)})
\end{align*}
and that
\begin{align*}
&H(\mathsf{U}^{(\ell_1+1,m+1)}|\mathsf{M}_{[\ell+1:m]})+H(\mathsf{M}_{[\ell+1:m]})\\
&\hspace{40pt}=H(\mathsf{U}^{(\ell_1+1,m+1)},\mathsf{M}_{[\ell+1:m]})\stackrel{(b)}{=}H(\mathsf{U}^{(\ell_1+1,m+1)})
\end{align*}
complete the proof of \eqref{eq:JH}. Here, $(a)$ and $(b)$ are due to the facts that $\mathsf{M}_{[\ell+1:m]}$ is a function of $\mathsf{W}_{[1:m]}$, which is a function of both $\mathsf{U}^{(\ell_1+1,m)}$ and $\mathsf{U}^{(\ell_1+1,m+1)}$ by Lemma~\ref{lemma1}.

To see \eqref{eq:prop3}, add \eqref{eq:JH} over $m\in[\ell+1:d-1]$ and cancel $\sum_{m=\ell+2}^{d-1}H(\mathsf{U}^{(\ell_1+1,m)})$ from both sides of the inequality. We have
\begin{align}
& H(\mathsf{U}^{(\ell_1+1,\ell+1)})+\frac{T_{d,d,\ell+1}}{d-\ell}H(\mathsf{U}^{(\ell_1,\ell+1)})\nonumber\\
& \hspace{40pt} \geq \sum_{m=\ell+1}^{d-1}\left((d-m)\sum_{j=\ell+1}^{m}T_{d,j,\ell}^{-1}B_j\right)+\nonumber\\
& \hspace{80pt} H(\mathsf{U}^{(\ell_1+1,d)})+\frac{T_{d,d,\ell+1}}{d-\ell}H(\mathsf{U}^{(\ell_1,\ell)}).\label{eq:JH1}
\end{align}
Note that
\begin{align}
& \sum_{m=\ell+1}^{d-1}\left((d-m)\sum_{j=\ell+1}^{m}T_{d,j,\ell}^{-1}B_j\right)\nonumber\\
& \hspace{10pt} = \sum_{j=\ell+1}^{d-1}T_{d,j,\ell}^{-1}B_j\left(\sum_{m=j}^{d-1}(d-m)\right)
= \sum_{j=\ell+1}^{d-1}T_{d,j,\ell}^{-1}T_{d,d,j}B_j.\label{eq:JH2}
\end{align}
Furthermore,
\begin{align}
H(\mathsf{U}^{(\ell_1+1,d)})& \stackrel{(a)}{=} H(\mathsf{U}^{(\ell_1+1,d)},\mathsf{M}_{[\ell+1:d]})\nonumber\\
& \stackrel{(b)}{=} H(\mathsf{U}^{(\ell_1+1,d)}|\mathsf{M}_{[\ell+1:d]})+H(\mathsf{M}_{[\ell+1:d]})\nonumber\\
& \stackrel{(c)}{=} H(\mathsf{U}^{(\ell_1+1,d)}|\mathsf{M}_{[\ell+1:d]})+\sum_{j=\ell+1}^dB_j\nonumber\\
& \geq H(\mathsf{U}^{(\ell_1,\ell)}|\mathsf{M}_{[\ell+1:d]})+\sum_{j=\ell+1}^dB_j\nonumber\\
& \stackrel{(d)}{=} H(\mathsf{U}^{(\ell_1,\ell)})+\sum_{j=\ell+1}^dB_j\label{eq:JH3}
\end{align}
where $(a)$ follows from the fact that $\mathsf{M}_{[\ell+1:d]}$ is a function of $\mathsf{W}_{[1:d]}$, which is is a function of $\mathsf{U}^{(\ell_1+1,d)}$ by Lemma~\ref{lemma1}; $(b)$ follows from the chain rule for entropy; $(c)$ follows from the fact that
$H(\mathsf{M}_{[\ell+1:d]})=\sum_{j=\ell+1}^dB_j$; and $(d)$ follows from the fact that $I(\mathsf{U}^{(\ell_1,\ell)};\mathsf{M}_{[\ell+1:d]})=0$ due to the repair secrecy requirement \eqref{eq:RepairSecrecy}.

Substituting \eqref{eq:JH2} and \eqref{eq:JH3} into \eqref{eq:JH1} gives:
\begin{align*}
& H(\mathsf{U}^{(\ell_1+1,\ell+1)})+\frac{T_{d,d,\ell+1}}{d-\ell}H(\mathsf{U}^{(\ell_1,\ell+1)})\\
& \geq \sum_{j=\ell+1}^{d-1}T_{d,j,\ell}^{-1}T_{d,d,j}B_j+\sum_{j=\ell+1}^dB_j+\\
& \hspace{20pt} \left(1+\frac{T_{d,d,\ell+1}}{d-\ell}\right)H(\mathsf{U}^{(\ell_1,\ell)})\\
& = \sum_{j=\ell+1}^{d-1}T_{d,j,\ell}^{-1}(T_{d,d,j}+T_{d,j,\ell})B_j+B_d+\frac{T_{d,d,\ell}}{d-\ell}H(\mathsf{U}^{(\ell_1,\ell)})\\
& \stackrel{(a)}{=} T_{d,d,\ell}\sum_{j=\ell+1}^{d-1}T_{d,j,\ell}^{-1}B_j+B_d+\frac{T_{d,d,\ell}}{d-\ell}H(\mathsf{U}^{(\ell_1,\ell)})\\
& = T_{d,d,\ell}\sum_{j=\ell+1}^{d}T_{d,j,\ell}^{-1}B_j+\frac{T_{d,d,\ell}}{d-\ell}H(\mathsf{U}^{(\ell_1,\ell)})
\end{align*}
where $(a)$ follows from the fact that $T_{d,d,j}+T_{d,j,\ell}=T_{d,d,\ell}$. This completes the proof of the proposition.
\end{IEEEproof}

We are now ready to prove the outer bounds \eqref{eq:B3} and \eqref{eq:B4}. To prove \eqref{eq:B3}, note that
\begin{align*}
\beta+\frac{1}{d-\ell}H(\mathsf{U}^{(\ell_1,\ell)}) & \stackrel{(a)}{\geq} \frac{1}{d-\ell}\left(H(\overline{\mathsf{S}}_{\rightarrow \ell+1})+H(\mathsf{U}^{(\ell_1,\ell)})\right)\\
& \stackrel{(b)}{\geq} \frac{1}{d-\ell}H(\mathsf{U}^{(\ell_1,\ell+1)})\\
& \stackrel{(c)}{\geq} \sum_{j=\ell+1}^{d}T_{d,j,\ell}^{-1}B_j+\frac{1}{d-\ell}H(\mathsf{U}^{(\ell_1,\ell)})
\end{align*}
where $(a)$ follows from the fact that $H(\overline{\mathsf{S}}_{\rightarrow \ell+1}) \leq (d-\ell)\beta$; $(b)$ follows from the union bound on entropy; and $(c)$ follows from \eqref{eq:prop1} of Proposition~\ref{prop1}. Cancelling $\frac{1}{d-\ell}H(\mathsf{U}^{(\ell_1,\ell)})$ from both sides of the inequality and normalizing both sides by $\sum_{t=\ell+1}^{d}B_t$ complete the proof of \eqref{eq:B3}.

To prove \eqref{eq:B4}, note that
\begin{align*}
\alpha&+T_{d,d,\ell_1+1}\beta+\frac{\ell_1+T_{d,d,\ell_1}}{d-\ell}H(\mathsf{U}^{(\ell_1,\ell)})\\
&\stackrel{(a)}{=} \alpha+T_{d,d,\ell_1+1}\beta+\\
& \hspace{10pt}\left(\frac{\ell_1}{d-\ell}+\frac{T_{d,\ell,\ell_1+1}}{d-\ell}
+\frac{T_{d,d,\ell+1}}{d-\ell}+\frac{d-\ell_1}{d-\ell}+1\right)H(\mathsf{U}^{(\ell_1,\ell)})\\
&=\frac{T_{d,\ell,\ell_1+1}}{d-\ell}H(\mathsf{U}^{(\ell_1,\ell)})+H(\mathsf{U}^{(\ell_1,\ell)})+\\
& \hspace{20pt}\alpha+T_{d,d,\ell_1+1}\beta+\\
& \hspace{20pt} \left(\frac{\ell_1}{d-\ell}+\frac{T_{d,d,\ell+1}}{d-\ell}+\frac{d-\ell_1}{d-\ell}\right)H(\mathsf{U}^{(\ell_1,\ell)})\\
&\stackrel{(b)}{\geq} \frac{T_{d,\ell,\ell_1+1}}{d-\ell}H(\mathsf{U}^{(\ell_1,\ell+1)})+H(\mathsf{U}^{(\ell_1,\ell_1+1)})+\\
& \hspace{20pt}\alpha+T_{d,d,\ell_1+1}\beta+\\
& \hspace{20pt}
\left(\frac{\ell_1}{d-\ell}+\frac{T_{d,d,\ell+1}}{d-\ell}+\frac{d-\ell_1}{d-\ell}\right)H(\mathsf{U}^{(\ell_1,\ell)})\\
&= \frac{d-\ell_1}{d-\ell}H(\mathsf{U}^{(\ell_1,\ell)})+H(\mathsf{U}^{(\ell_1,\ell_1+1)},\mathsf{S}_{\ell_1+1\rightarrow [1:\ell_1]})+\\
& \hspace{20pt}\alpha+T_{d,d,\ell_1+1}\beta+\left(\frac{\ell_1}{d-\ell}
+\frac{T_{d,d,\ell+1}}{d-\ell}\right)H(\mathsf{U}^{(\ell_1,\ell)})+\\
& \hspace{20pt} \frac{T_{d,\ell,\ell_1+1}}{d-\ell}H(\mathsf{U}^{(\ell_1,\ell+1)})\\
&\stackrel{(c)}{\geq} \frac{d-\ell_1}{d-\ell}H(\mathsf{U}^{(\ell_1,\ell+1)})+H(\mathsf{U}^{(\ell_1,\ell_1)},\mathsf{S}_{\ell_1+1\rightarrow [1:\ell_1]})+\\
& \hspace{20pt}\alpha+T_{d,d,\ell_1+1}\beta+\left(\frac{\ell_1}{d-\ell}
+\frac{T_{d,d,\ell+1}}{d-\ell}\right)H(\mathsf{U}^{(\ell_1,\ell)})+\\
& \hspace{20pt} \frac{T_{d,\ell,\ell_1+1}}{d-\ell}H(\mathsf{U}^{(\ell_1,\ell+1)})\\
&\stackrel{(d)}{=} \alpha+H(\mathsf{U}^{(\ell_1,\ell_1)},\mathsf{S}_{\ell_1+1\rightarrow [1:\ell_1]})+\\
& \hspace{20pt}T_{d,d,\ell_1+1}\beta+\left(\frac{\ell_1}{d-\ell}
+\frac{T_{d,d,\ell+1}}{d-\ell}\right)H(\mathsf{U}^{(\ell_1,\ell)})+\\
& \hspace{20pt} \frac{T_{d,\ell,\ell_1}}{d-\ell}H(\mathsf{U}^{(\ell_1,\ell+1)})\\
&\stackrel{(e)}{\geq} H(W_{\ell+1})+H(\mathsf{U}^{(\ell_1,\ell_1)},\mathsf{S}_{\ell_1+1\rightarrow [1:\ell_1]})+\\
& \hspace{20pt}T_{d,d,\ell_1+1}\beta+\left(\frac{\ell_1}{d-\ell}
+\frac{T_{d,d,\ell+1}}{d-\ell}\right)H(\mathsf{U}^{(\ell_1,\ell)})+\\
& \hspace{20pt} \frac{T_{d,\ell,\ell_1}}{d-\ell}H(\mathsf{U}^{(\ell_1,\ell+1)})\\
&\stackrel{(f)}{=} H(W_{\ell+1},\mathsf{S}_{\ell_1+1\rightarrow [1:\ell_1]})+H(\mathsf{U}^{(\ell_1,\ell_1)},\mathsf{S}_{\ell_1+1\rightarrow [1:\ell_1]})+\\
& \hspace{20pt}T_{d,d,\ell_1+1}\beta+\left(\frac{\ell_1}{d-\ell}
+\frac{T_{d,d,\ell+1}}{d-\ell}\right)H(\mathsf{U}^{(\ell_1,\ell)})+\\
& \hspace{20pt} \frac{T_{d,\ell,\ell_1}}{d-\ell}H(\mathsf{U}^{(\ell_1,\ell+1)})\\
&\stackrel{(g)}{\geq} H(\mathsf{S}_{\ell_1+1\rightarrow [1:\ell_1]})+H(\mathsf{U}^{(\ell_1+1,\ell_1+1)},\mathsf{S}_{\ell_1+1\rightarrow [1:\ell_1]})+\\
& \hspace{20pt}T_{d,d,\ell_1+1}\beta+\left(\frac{\ell_1}{d-\ell}
+\frac{T_{d,d,\ell+1}}{d-\ell}\right)H(\mathsf{U}^{(\ell_1,\ell)})+\\
& \hspace{20pt} \frac{T_{d,\ell,\ell_1}}{d-\ell}H(\mathsf{U}^{(\ell_1,\ell+1)})\\
&\stackrel{(h)}{\geq} H(\mathsf{S}_{\ell_1+1\rightarrow [1:\ell_1]})+\frac{\ell_1}{d-\ell}H(\mathsf{U}^{(\ell_1,\ell)})+\\
& \hspace{20pt}H(\mathsf{U}^{(\ell_1+1,\ell_1+1)})+T_{d,d,\ell_1+1}\beta
+\frac{T_{d,d,\ell+1}}{d-\ell}H(\mathsf{U}^{(\ell_1,\ell)})+\\
& \hspace{20pt}\frac{T_{d,\ell,\ell_1}}{d-\ell}H(\mathsf{U}^{(\ell_1,\ell+1)})\\
&\stackrel{(i)}{\geq} \frac{\ell_1}{d-\ell}H(\mathsf{U}^{(\ell_1,\ell+1)})+
H(\mathsf{U}^{(\ell_1+1,\ell_1+1)})+T_{d,d,\ell_1+1}\beta\\
& \hspace{20pt}
+\frac{T_{d,d,\ell+1}}{d-\ell}H(\mathsf{U}^{(\ell_1,\ell)})
+\frac{T_{d,\ell,\ell_1}}{d-\ell}H(\mathsf{U}^{(\ell_1,\ell+1)})\\
&\stackrel{(j)}{=} H(\mathsf{U}^{(\ell_1+1,\ell_1+1)})+T_{d,\ell+1,\ell_1+1}\beta\\
& \hspace{20pt}
+\frac{T_{d,d,\ell+1}}{d-\ell}H(\mathsf{U}^{(\ell_1,\ell)})+T_{d,d,\ell+1}\beta+\\
& \hspace{20pt}
\frac{T_{d,\ell,\ell_1}+\ell_1}{d-\ell}H(\mathsf{U}^{(\ell_1,\ell+1)})\\
&\stackrel{(k)}{\geq} H(\mathsf{U}^{(\ell_1+1,\ell+1)})+\frac{T_{d,d,\ell+1}}{d-\ell}H(\mathsf{U}^{(\ell_1,\ell+1)})+\\
& \hspace{20pt}\frac{T_{d,\ell,\ell_1}+\ell_1}{d-\ell}H(\mathsf{U}^{(\ell_1,\ell+1)})\\
&\stackrel{(l)}{\geq} (T_{d,d,\ell}+T_{d,\ell,\ell_1}+\ell_1)\sum_{j=\ell+1}^{d}T_{d,j,\ell}^{-1}B_j+\\
& \hspace{20pt} \frac{T_{d,d,\ell}+T_{d,\ell,\ell_1}+\ell_1}{d-\ell}H(\mathsf{U}^{(\ell_1,\ell)})\\
&\stackrel{(m)}{=} (T_{d,d,\ell_1}+\ell_1)\sum_{j=\ell+1}^{d}T_{d,j,\ell}^{-1}B_j
+\frac{T_{d,d,\ell_1}+\ell_1}{d-\ell}H(\mathsf{U}^{(\ell_1,\ell)})
\end{align*}
where $(a)$ follows from the fact that $T_{d,\ell,\ell_1+1}+T_{d,d,\ell+1}+d-\ell_1+d-\ell=T_{d,d,\ell_1} $; $(b)$ follows from Corollary \ref{coro1} that
\begin{align*}
&\frac{T_{d,\ell,\ell_1+1}}{d-\ell}H(\mathsf{U}^{(\ell_1,\ell)})+H(\mathsf{U}^{(\ell_1,\ell)})\geq \\
 &\hspace{40pt}\frac{T_{d,\ell,\ell_1+1}}{d-\ell}H(\mathsf{U}^{(\ell_1,\ell+1)})+H(\mathsf{U}^{(\ell_1,\ell_1+1)})
\end{align*}
by setting $j_1=\ell$, $j_2=\ell_1+1$ and $i=i'=\ell_1$ in \ref{eq:coro1}; $(c)$ follows from \eqref{eq:mutant exchange} in Lemma \ref{lemma:mutant exchange}; $(d)$ follows from the fact that $T_{d,\ell,\ell_1+1}+d-\ell_1=T_{d,\ell,\ell_1} $; $(e)$ follows from the fact that $H(\mathsf{W}_{\ell+1}) \leq \alpha$; $(f)$ and $(h)$ follows from the fact that $\mathsf{S}_{\ell_1+1\rightarrow [1:\ell_1]}$ is a function of $\mathsf{W}_{\ell_1+1}$; $(g)$ follows from the fact that $H(W_{\ell+1},\mathsf{S}_{\ell_1+1\rightarrow [1:\ell_1]})+H(\mathsf{U}^{(\ell_1,\ell_1)},\mathsf{S}_{\ell_1+1\rightarrow [1:\ell_1]})\geq H(\mathsf{S}_{\ell_1+1\rightarrow [1:\ell_1]})+H(\mathsf{U}^{(\ell_1+1,\ell_1+1)},\mathsf{S}_{\ell_1+1\rightarrow [1:\ell_1]})$ due to submodularity; $(i)$ follows from \eqref{eq:prop2} in Proposition \ref{prop2}; $(j)$ follows from the fact that $T_{d,d,\ell_1+1}=T_{d,d,\ell+1}+T_{d,\ell+1,\ell_1+1}$; $(k)$ follows from the facts that $T_{d,\ell+1,\ell_1+1}\beta \geq \overline{\mathsf{S}}_{\rightarrow [\ell_1+2:\ell+1]}$ and $(d-\ell)\beta \geq \overline{\mathsf{S}}_{\rightarrow \ell+1}$; $(l)$ follows from \eqref{eq:prop1} and \eqref{eq:prop3} of Proposition
\ref{prop1} and \ref{prop3} respectively; $(m)$ follows from the fact that $T_{d,d,\ell}+T_{d,\ell,\ell_1}=T_{d,d,\ell_1}$. Cancelling $\frac{T_{d,d,\ell_1}+\ell_1}{d-\ell}H(\mathsf{U}^{(\ell_1,\ell)})$ from both sides of the inequality and normalizing both sides by $\sum_{t=\ell+1}^{d}B_t$ complete the proof of \eqref{eq:B4}.

\section{Concluding remarks}\label{sec:Con}
This paper considered the problem of MDC-SR with a generalized eavesdropping model. It was shown that the MBR point of the achievable normalized storage-capacity repair-bandwidth tradeoff region depends on the numbers of type \uppercase\expandafter{\romannumeral1} and type \uppercase\expandafter{\romannumeral2} compromised storage nodes only via their total, as long as the number of type \uppercase\expandafter{\romannumeral1} compromised nodes is less than or equal to the number of type \uppercase\expandafter{\romannumeral2} compromised nodes. Moving forward, it would be interesting to see whether this result extends to the entire achievable normalized storage-capacity repair-bandwidth tradeoff region.

\appendix[Proof of the Exchange Lemma \uppercase\expandafter{\romannumeral2}]

First note that $d-\ell_1 > d-\ell$, so we may write $d-\ell_1=s(d-\ell)+r$ for some integer $s\geq 1$ and $r\in[1:d-\ell]$. Next, let
$$ a_t:=\left\{
\begin{array}{lcl}
t ,     &      & {t     \in   [1:\ell_1 ]   }\\
t+\ell_1,   &      & {t \in [\ell_1+1: \ell-\ell_1]}\\
t+j ,      &      & {t \in [\ell-\ell_1+1: d-\ell]}.
\end{array} \right. $$
Finally, let $\tau_0:=\{a_t: t\in [1: r]\}$ and
\begin{align*}
\tau_q:=\{a_t: t\in [r+1+(q-1)(d-\ell):r+q(d-\ell)]\}
\end{align*}
for any $q\in[1:s]$. It is straightforward to verify that:

\begin{itemize}
\item $\tau_q  \cap  \tau_{q'} =\emptyset$ for any $q\neq q'$;
\item $\bigcup_{q=0}^{s-1}\tau_q =[1:\ell_1]\cup [2\ell_1+1:\ell]$;
\item $\tau_s=[\ell+2:d+1]$.
\end{itemize}

Consider a symmetrical $(n=d+1,d,N_1,\ldots,N_d,T,S)$ code that satisfies the node regeneration requirement \eqref{eq:NodeRegen}. Let us show by induction that for any $p\in[1:s]$, we have
\begin{align}
&pH(\mathsf{U}^{(\ell_1,\ell)}|\mathsf{M}^{(\ell)})+H(\mathsf{U}^{(\ell_1,\ell_1+1)}|\mathsf{M}^{(\ell)})\nonumber\\
& \geq pH(\mathsf{U}^{(\ell_1,\ell+1)}|\mathsf{M}^{(\ell)})+\nonumber\\
& \hspace{20pt} H(\mathsf{W}_{[\ell_1+1:2\ell_1]},\mathsf{S}_{\ell_1 \rightarrow [ \ell_1+1:2\ell_1]},\mathsf{S}_{\bigcup_{q=0}^{s-p}\tau_q\rightarrow \ell+1}|\mathsf{M}^{(\ell)}).\label{eq:QQ1}
\end{align}

To prove the base case of $p=1$, first note that
\begin{align*}
H&(\mathsf{U}^{(\ell_1,\ell)}|\mathsf{M}^{(\ell)})\\
&\stackrel{(a)}{=} H(\mathsf{U}^{(\ell_1,\ell)},\mathsf{W}_{[1:\ell_1]},\underline{\mathsf{S}}_{\rightarrow [\ell_1+1:\ell]}|\mathsf{M}^{(\ell)})\\
&= H(\mathsf{W}_{[1:\ell_1]},\mathsf{S}_{\rightarrow [\ell_1+1:\ell]}|\mathsf{M}^{(\ell)})\\
&\stackrel{(b)}{=} H(\mathsf{W}_{[1:\ell]},\mathsf{S}_{\rightarrow [\ell_1+1:\ell]},\mathsf{S}_{[1:\ell]\rightarrow \ell+1}|\mathsf{M}^{(\ell)})
\end{align*}
where $(a)$ follows from the fact that $(\mathsf{W}_{[\ell_1:\ell]},\underline{\mathsf{S}}_{\rightarrow [\ell_1+1:\ell]})$ is a function of $\mathsf{U}^{(\ell_1,\ell)}$ by Lemma~\ref{lemma1}, and $(b)$ follows from the fact that $\mathsf{S}_{[1:\ell]\rightarrow \ell+1}$ is a function of $\mathsf{W}_{[1:\ell]}$. Furthermore,
\begin{align*}
H&(\mathsf{U}^{(\ell_1,\ell_1+1)}|\mathsf{M}^{(\ell)})\\
&\stackrel{(a)}{=} H(\mathsf{U}^{(\ell_1,\ell_1+1)},\underline{\mathsf{S}}_{\rightarrow \ell_1+1}|\mathsf{M}^{(\ell)})\\
&= H(\mathsf{W}_{[1:\ell_1]},\mathsf{S}_{\rightarrow \ell_1+1}|\mathsf{M}^{(\ell)})\\
&\stackrel{(b)}{=} H(\mathsf{W}_{[\ell_1+1:2\ell_1]},\mathsf{S}_{\rightarrow 2\ell_1+1}|\mathsf{M}^{(\ell)})\\
&\stackrel{(c)}{=} H(\mathsf{W}_{[\ell_1+1:2\ell_1]},\mathsf{S}_{\rightarrow \ell+1}|\mathsf{M}^{(\ell)})\\
&\stackrel{(d)}{=} H(\mathsf{W}_{[\ell_1+1:2\ell_1]},\mathsf{S}_{[1:\ell_1]\rightarrow \ell+1},\mathsf{S}_{[2\ell_1+1:\ell]\rightarrow \ell+1},\\
&\quad\mathsf{S}_{[\ell+2:d+1]\rightarrow \ell+1},\mathsf{S}_{\ell+1\rightarrow [\ell_1+1:2\ell_1]}|\mathsf{M}^{(\ell)})
\end{align*}
where $(a)$ follows from the fact that $\underline{\mathsf{S}}_{\rightarrow\ell_1+1}$ is a function of $\mathsf{U}^{(\ell_1,\ell_1+1)}$ by Lemma~\ref{lemma1}, and $(b)$ and $(c)$ follow from the symmetrical code that we consider; $(d)$ follows that $\mathsf{S}_{\ell+1\rightarrow [\ell_1+1:2\ell_1]}$ is a function of $\mathsf{S}_{\rightarrow \ell+1}$ . It follows that
\begin{align*}
&H(\mathsf{U}^{(\ell_1,\ell)}|\mathsf{M}^{(\ell)})+H(\mathsf{U}^{\ell_1,\ell_1+1}|\mathsf{M}^{(\ell)})\\
&\geq H(\mathsf{W}_{[1:\ell]},\mathsf{S}_{\rightarrow [\ell_1+1:\ell]},\mathsf{S}_{[1:\ell]\rightarrow \ell+1}|\mathsf{M}^{(\ell)})+\\
& \hspace{20pt} H(\mathsf{W}_{[\ell_1+1:2\ell_1]},\mathsf{S}_{[1:\ell_1]\rightarrow \ell+1},\mathsf{S}_{[2\ell_1:\ell]\rightarrow \ell+1},\\
& \hspace{20pt} \mathsf{S}_{[\ell+2:d+1]\rightarrow \ell+1}, \mathsf{S}_{\ell+1\rightarrow [\ell_1+1:2\ell_1]}|\mathsf{M}^{(\ell)})\\
& \stackrel{(a)}{\geq}H(\mathsf{W}_{[1:\ell]},\mathsf{S}_{\rightarrow [\ell_1+1:\ell]},\\
& \hspace{20pt} \mathsf{S}_{[1:\ell]\rightarrow \ell+1},\mathsf{S}_{[\ell+2:d+1]\rightarrow \ell+1}|\mathsf{M}^{(\ell)})+\\
& \hspace{20pt} H(\mathsf{W}_{[\ell_1+1:2\ell_1]},\mathsf{S}_{[1:\ell_1]\rightarrow \ell+1},\mathsf{S}_{[2\ell_1:\ell]\rightarrow \ell+1},\\
& \hspace{20pt} \mathsf{S}_{\ell+1\rightarrow [\ell_1+1:2\ell_1]}|\mathsf{M}^{(\ell)})\\
& = H(\mathsf{U}^{(\ell_1,\ell)},\underline{\mathsf{S}}_{\rightarrow \ell+1}|\mathsf{M}^{(\ell)})+\\
& \hspace{20pt} H(\mathsf{W}_{[\ell_1+1:2\ell_1]},\mathsf{S}_{\ell+1\rightarrow [\ell_1+1:2\ell_1]},\mathsf{S}_{\bigcup_{q=0}^{s-1}\tau_q\rightarrow \ell+1}|\mathsf{M}^{(\ell)})\\
& = H(\mathsf{U}^{(\ell_1,\ell+1)}|\mathsf{M}^{(\ell)})+\\
& \hspace{20pt} H(\mathsf{W}_{[\ell_1+1:2\ell_1]},\mathsf{S}_{\ell+1\rightarrow [\ell_1+1:2\ell_1]},\mathsf{S}_{\bigcup_{q=0}^{s-1}\tau_q\rightarrow \ell+1}|\mathsf{M}^{(\ell)})
\end{align*}
where $(a)$ follows from the submodularity of the entropy function. This completes the proof of the base case of $p=1$.

Assume that \eqref{eq:QQ1} holds for some $p\in[1:s-1]$. We have
\begin{align}
&(p+1)H(\mathsf{U}^{(\ell_1,\ell)}|\mathsf{M}^{(\ell)})+H(\mathsf{U}^{(\ell_1,\ell_1+1)}|\mathsf{M}^{(\ell)})\nonumber\\
& = H(\mathsf{U}^{(\ell_1,\ell)}|\mathsf{M}^{(\ell)})+\left(pH(\mathsf{U}^{(\ell_1,\ell)}|\mathsf{M}^{(\ell)})
+H(\mathsf{U}^{(\ell_1,\ell_1+1)}|\mathsf{M}^{(\ell)})\right)\nonumber\\
& \geq H(\mathsf{U}^{(\ell_1,\ell)}|\mathsf{M}^{(\ell)})+pH(\mathsf{U}^{(\ell_1,\ell+1)}|\mathsf{M}^{(\ell)})+\nonumber\\
& \hspace{20pt} H(\mathsf{W}_{[\ell_1+1:2\ell_1]},\mathsf{S}_{\ell+1 \rightarrow [\ell_1+1:2\ell_1]},\mathsf{S}_{\bigcup_{q=0}^{s-p}\tau_q\rightarrow \ell+1}|\mathsf{M}^{(\ell)}).\label{eq:QQ2}
\end{align}
Note that both $\mathsf{S}_{\ell+1 \rightarrow[\ell_1+1:2\ell_1]}$ and $\mathsf{S}_{\bigcup_{q=0}^{s-(p+1)}\tau_q\rightarrow \ell+1}$ are functions of $\mathsf{W}_{[1:\ell]}$, which is in turn a function of $\mathsf{U}^{(\ell_1,\ell)}$ by Lemma~\ref{lemma1}. We thus have
\begin{align*}
H&(\mathsf{U}^{(\ell_1,\ell)}|\mathsf{M}^{(\ell)})\\
&=H(\mathsf{U}^{(\ell_1,\ell)},\mathsf{S}_{\ell+1 \rightarrow[\ell_1+1:2\ell_1]}, \mathsf{S}_{\bigcup_{q=0}^{s-(p+1)}\tau_q\rightarrow \ell+1}|\mathsf{M}^{(\ell)}).
\end{align*}
Furthermore, by the symmetrical code that we consider we have
\begin{align*}
H&(\mathsf{W}_{[\ell_1+1:2\ell_1]},\mathsf{S}_{\ell+1 \rightarrow [\ell_1+1:2\ell_1]},\mathsf{S}_{\bigcup_{q=0}^{s-p}\tau_q\rightarrow \ell+1}|\mathsf{M}^{(\ell)})\\
& = H(\mathsf{W}_{[\ell_1+1:2\ell_1]},\mathsf{S}_{\ell+1 \rightarrow [\ell_1+1:2\ell_1]},\\
& \hspace{20pt} \mathsf{S}_{\bigcup_{q=0}^{s-(p+1)}\tau_q\rightarrow \ell+1},\mathsf{S}_{[\ell+2:d+1]\rightarrow \ell+1}|\mathsf{M}^{(\ell)}).
\end{align*}
It follows that
\begin{align}
&H(\mathsf{U}^{(\ell_1,\ell)}|\mathsf{M}^{(\ell)})+\nonumber\\
& \hspace{20pt} H(\mathsf{W}_{[\ell_1+1:2\ell_1]},\mathsf{S}_{\ell+1 \rightarrow [\ell_1+1:2\ell_1]},\mathsf{S}_{\bigcup_{q=0}^{s-p}\tau_q\rightarrow \ell+1}|\mathsf{M}^{(\ell)})\nonumber\\
& = H(\mathsf{U}^{(\ell_1,\ell)}, \mathsf{S}_{\ell+1 \rightarrow [\ell_1+1,2\ell_1]},\mathsf{S}_{\bigcup_{q=0}^{s-(p+1)}\tau_q\rightarrow \ell+1}|\mathsf{M}^{(\ell)})+\nonumber\\
& \hspace{20pt} H(\mathsf{W}_{[\ell_1+1:2\ell_1]},\mathsf{S}_{\ell+1 \rightarrow [\ell_1+1:2\ell_1]},\nonumber\\
& \hspace{20pt} \mathsf{S}_{\bigcup_{q=0}^{s-(p+1)}\tau_q\rightarrow \ell+1},\mathsf{S}_{[\ell+2:d+1]\rightarrow \ell+1}|\mathsf{M}^{(\ell)})\nonumber\\
& \stackrel{(a)}{\geq} H(\mathsf{U}^{(\ell_1,\ell)}, \mathsf{S}_{\ell+1 \rightarrow [\ell_1+1,2\ell_1]},\mathsf{S}_{\bigcup_{q=0}^{s-(p+1)}\tau_q\rightarrow \ell+1},\nonumber\\
& \hspace{20pt}  \mathsf{S}_{[\ell+2:d+1]\rightarrow \ell+1}|\mathsf{M}^{(\ell)})+\nonumber\\
& \hspace{20pt} H(\mathsf{W}_{[\ell_1+1:2\ell_1]},\mathsf{S}_{\ell+1 \rightarrow [\ell_1+1:2\ell_1]}, \mathsf{S}_{\bigcup_{q=0}^{s-(p+1)}\tau_q\rightarrow \ell+1}|\mathsf{M}^{(\ell)})\nonumber\\
& = H(\mathsf{U}^{(\ell_1,\ell+1)}|\mathsf{M}^{(\ell)})+\nonumber\\
& \hspace{20pt} H(\mathsf{W}_{[\ell_1+1:2\ell_1]},\mathsf{S}_{\ell+1 \rightarrow [\ell_1+1:2\ell_1]},\mathsf{S}_{\bigcup_{q=0}^{s-(p+1)}\tau_q\rightarrow \ell+1}|\mathsf{M}^{(m)})\label{eq:QQ3}
\end{align}
where $(a)$ follows from the submodularity of the entropy function. Substituting \eqref{eq:QQ3} into \eqref{eq:QQ2} gives
\begin{align*}
&(p+1)H(\mathsf{U}^{(\ell_1,\ell)}|\mathsf{M}^{(\ell)})+H(\mathsf{U}^{(\ell_1,\ell_1+1)}|\mathsf{M}^{(\ell)})\\
& \geq (p+1)H(\mathsf{U}^{(\ell_1,\ell+1)}|\mathsf{M}^{(\ell)})+\nonumber\\
& \hspace{20pt} H(\mathsf{W}_{[\ell_1+1:2\ell_1]},\mathsf{S}_{\ell+1\rightarrow [\ell_1+1:2\ell_1]},\mathsf{S}_{\bigcup_{q=0}^{s-(p+1)}\tau_q\rightarrow \ell+1}|\mathsf{M}^{(\ell)})
\end{align*}
which completes the induction step and hence the proof of \eqref{eq:QQ1}.

Setting $p=s$ in \eqref{eq:QQ1}, we have
\begin{align}
&sH(\mathsf{U}^{(\ell_1,\ell)}|\mathsf{M}^{(\ell)})+H(\mathsf{U}^{(\ell_1,\ell_1+1)}|\mathsf{M}^{(\ell)})\nonumber\\
& \geq sH(\mathsf{U}^{(\ell_1,\ell+1)}|\mathsf{M}^{(\ell)})+\nonumber\\
& \hspace{20pt} H(\mathsf{W}_{[\ell_1+1:2\ell_1]},\mathsf{S}_{\ell+1\rightarrow [\ell_1+1:2\ell_1]},\mathsf{S}_{\tau_0\rightarrow \ell+1}|\mathsf{M}^{(\ell)})\nonumber\\
& = sH(\mathsf{U}^{(\ell_1,\ell+1)}|\mathsf{M}^{(\ell)})+H(\mathsf{W}_{[\ell_1+1:2\ell_1]},\mathsf{S}_{\ell+1\rightarrow [\ell_1+1:2\ell_1]}|\mathsf{M}^{(\ell)})+\nonumber\\
& \hspace{20pt} H(\mathsf{S}_{\tau_0\rightarrow \ell+1}|\mathsf{W}_{[\ell_1+1:2\ell_1]},\mathsf{S}_{\ell+1\rightarrow [\ell_1+1:2\ell_1]},\mathsf{M}^{(\ell)}).\label{eq:QQ4}
\end{align}
By the symmetrical codes that we consider, we have
\begin{align}
H&(\mathsf{W}_{[\ell_1+1:2\ell_1]},\mathsf{S}_{\ell+1\rightarrow [\ell_1+1:2\ell_1]}|\mathsf{M}^{(\ell)})\nonumber\\
&=H(\mathsf{W}_{[1:\ell_1]},\mathsf{S}_{\ell+1\rightarrow [1:\ell_1]}|\mathsf{M}^{(\ell)})\nonumber\\
&=H(\mathsf{W}_{[1:\ell_1]},\mathsf{S}_{\ell_1+1\rightarrow [1:\ell_1]}|\mathsf{M}^{(\ell)})\label{eq:QQ5}
\end{align}
and
\begin{align*}
&H(\mathsf{S}_{\tau_0\rightarrow \ell+1}|\mathsf{W}_{[\ell_1+1:2\ell_1]},\mathsf{S}_{\ell+1\rightarrow [\ell_1+1:2\ell_1]},\mathsf{M}^{(\ell)})\\
&=H(\mathsf{S}_{\tau\rightarrow \ell+1}|\mathsf{W}_{[\ell_1+1:2\ell_1]},\mathsf{S}_{\ell+1\rightarrow [\ell_1+1:2\ell_1]},\mathsf{M}^{(\ell)})
\end{align*}
for {\em any} subset $\tau \subseteq [\ell+2:d+1]$ such that $|\tau|=r$. By Han's subset inequality \cite{Han-IC78}, we have
\begin{align}
H&(\mathsf{S}_{\tau_0\rightarrow \ell+1}|\mathsf{W}_{[\ell_1+1:2\ell_1]},\mathsf{S}_{\ell+1\rightarrow [\ell_1+1:2\ell_1]},\mathsf{M}^{(\ell)})\nonumber\\
& \geq \frac{r}{d-\ell}H(\mathsf{S}_{[\ell+2:d+1]\rightarrow \ell+1}|\mathsf{W}_{[\ell_1+1:2\ell_1]},
\nonumber\\
& \hspace{20pt} \mathsf{S}_{\ell+1\rightarrow [\ell_1+1:2\ell_1]},\mathsf{M}^{(\ell)})\nonumber\\
& \geq \frac{r}{d-\ell}H(\mathsf{S}_{[\ell+2:d+1]\rightarrow \ell+1}|\mathsf{W}_{[\ell_1+1:2\ell_1]},
\nonumber\\
& \hspace{20pt} \mathsf{S}_{\ell+1\rightarrow [\ell_1+1:2\ell_1]},\mathsf{M}^{(\ell)},\mathsf{U}^{(\ell_1,\ell)})\nonumber\\
& \stackrel{(a)}{=} \frac{r}{d-\ell}H(\mathsf{S}_{[\ell+2:d+1]\rightarrow \ell+1}|\mathsf{U}^{(\ell_1,\ell)},\mathsf{M}^{(\ell)})\nonumber\\
& = \frac{r}{d-\ell}\left(H(\mathsf{S}_{[\ell+2:d+1]\rightarrow \ell+1},\mathsf{U}^{(\ell_1,\ell)}|\mathsf{M}^{(\ell)})-\right.\nonumber\\
& \hspace{20pt} \left.H(\mathsf{U}^{(\ell_1,\ell)}|\mathsf{M}^{(\ell)})\right)\nonumber\\
& = \frac{r}{d-\ell}\left(H(\mathsf{U}^{(\ell_1,\ell+1)}|\mathsf{M}^{(\ell)})-H(\mathsf{U}^{(\ell_1,\ell)}|\mathsf{M}^{(\ell)})\right)\label{eq:QQ6}
\end{align}
where $(a)$ follows from the fact that $(\mathsf{W}_{[\ell_1+1:2\ell_1]}, \mathsf{S}_{\ell+1\rightarrow [\ell_1+1:2\ell_1]})$ is a function of $\mathsf{U}^{(\ell_1,\ell)}$ by Lemma~\ref{lemma1}. Substituting \eqref{eq:QQ5} and \eqref{eq:QQ6} into \eqref{eq:QQ4} gives:
\begin{align*}
&\left(s+\frac{r}{d-\ell}\right)H(\mathsf{U}^{(\ell_1,\ell)}|\mathsf{M}^{(\ell)})+H(\mathsf{U}^{(\ell_1,\ell_1+1)}|\mathsf{M}^{(\ell)})\\
& \geq \left(s+\frac{r}{d-\ell}\right)H(\mathsf{U}^{(\ell_1,\ell+1)}|\mathsf{M}^{(\ell)})+H(\mathsf{U}^{(\ell_1,\ell_1)}|\mathsf{M}^{(\ell)})
\end{align*}
which is equivalent to \eqref{eq:exchange} by noting that
\begin{align*}
s+\frac{r}{d-\ell}=\frac{s(d-\ell)+r}{d-\ell}=\frac{d-\ell_1}{d-\ell}.
\end{align*}
This completes the proof of the exchange lemma.

\bibliographystyle{ieeetr}

\end{document}